\documentclass[dvips]{acta}
\usepackage{supertabular,lscape,epsfig}
\usepackage{amssymb}
\usepackage{amsmath}
\DeclareSymbolFont{ppa}{OT1}{ppl}{m}{it}
\DeclareMathSymbol{\vv}{\mathalpha}{ppa}{'166}

\newfont{\hb}{rphvb at 10pt}
\newfont{\hbo}{rphvbo at 10pt}
\newfont{\bitt}{rptmbi at 12pt}
\newfont{\bits}{rptmbi at 11pt}

\SetPages{1}{1}

\SetVol{65}{2015}

\begin{document}

\newcommand{\TabCapp}[2]{\begin{center}\parbox[t]{#1}{\centerline{
  \small {\spaceskip 2pt plus 1pt minus 1pt T a b l e}
  \refstepcounter{table}\thetable}
  \vskip2mm
  \centerline{\footnotesize #2}}
  \vskip3mm
\end{center}}

\newcommand{\TTabCap}[3]{\begin{center}\parbox[t]{#1}{\centerline{
  \small {\spaceskip 2pt plus 1pt minus 1pt T a b l e}
  \refstepcounter{table}\thetable}
  \vskip2mm
  \centerline{\footnotesize #2}
  \centerline{\footnotesize #3}}
  \vskip1mm
\end{center}}

\newcommand{\MakeTableSepp}[4]{\begin{table}[p]\TabCapp{#2}{#3}
  \begin{center} \TableFont \begin{tabular}{#1} #4
  \end{tabular}\end{center}\end{table}}

\newcommand{\MakeTableee}[4]{\begin{table}[htb]\TabCapp{#2}{#3}
  \begin{center} \TableFont \begin{tabular}{#1} #4
  \end{tabular}\end{center}\end{table}}

\newcommand{\MakeTablee}[5]{\begin{table}[htb]\TTabCap{#2}{#3}{#4}
  \begin{center} \TableFont \begin{tabular}{#1} #5
  \end{tabular}\end{center}\end{table}}

\newfont{\bb}{ptmbi8t at 12pt}
\newfont{\bbb}{cmbxti10}
\newfont{\bbbb}{cmbxti10 at 9pt}
\newcommand{\uprule}{\rule{0pt}{2.5ex}}
\newcommand{\douprule}{\rule[-2ex]{0pt}{4.5ex}}
\newcommand{\dorule}{\rule[-2ex]{0pt}{2ex}}
\def\thefootnote{\fnsymbol{footnote}}
\begin{Titlepage}
\Title{Eclipsing Binaries with Classical Cepheid Component\\
in the Magellanic System\footnote{Based on observations obtained with the 1.3-m Warsaw
telescope at the Las Campanas Observatory of the Carnegie Institution
for Science.}}
\Author{A.~~U~d~a~l~s~k~i$^1$,~~
I.~~S~o~s~z~y~ñ~s~k~i$^1$,~~
M.\,K.~~S~z~y~m~a~ñ~s~k~i$^1$,~~
G.~~P~i~e~t~r~z~y~ñ~s~k~i$^1$,\\
R.~~P~o~l~e~s~k~i$^{1,2}$,~~
P.~~P~i~e~t~r~u~k~o~w~i~c~z$^1$,~~
S.~~K~o~z~³~o~w~s~k~i$^1$,~~
P.~~M~r~ó~z$^1$\\
D.~~S~k~o~w~r~o~n$^1$,~~
J.~~S~k~o~w~r~o~n$^1$,~~
\L.~~W~y~r~z~y~k~o~w~s~k~i$^1$,\\
K.~~U~l~a~c~z~y~k$^3$~~
and~~M.~~P~a~w~l~a~k$^1$}
{$^1$Warsaw University Observatory, Al.~Ujazdowskie~4, 00-478~Warszawa, Poland\\
e-mail: (udalski,soszynsk)@astrouw.edu.pl\\
$^2$ Department of Astronomy, Ohio State University, 140 W. 18th Ave., Columbus, OH 43210, USA\\
$^3$ Department of Physics, University of Warwick, Gibbet Hill Road, Coventry, CV4 7AL, UK}
\end{Titlepage}

\Abstract{We present a census of eclipsing binary systems with
classical Cepheid as a component. All such systems known were found in the
OGLE collection of classical Cepheids in the Magellanic System. We
extend the list of potential candidates adding four new objects found in
the OGLE-IV photometric data.

One of the new Cepheids in the eclipsing system, OGLE-SMC-CEP-3235,
revealed only one eclipse during 15 years of the OGLE photometric
monitoring. However, it additionally shows very well pronounced
light-time effect indicating that the binarity is real. We also search
for the light-time effect in other known eclipsing Cepheids and we
clearly detect it in OGLE-LMC-CEP-1812. We discuss application of this
tool for the search for Cepheids in non-eclipsing binary systems.}
{Stars: variables: Cepheids -- binaries: eclipsing -- Magellanic Clouds}

\Section{Introduction}

Classical Cepheids form one of the most interesting groups of variable
stars. These pulsating objects reveal very sharp period--luminosity (PL)
relation, discovered yet at the beginning of the 20th century (Leavitt
1908). This feature makes classical Cepheids very useful standard
candles.  They serve as primary distance indicators in the nearby
Universe reaching as far as the Virgo cluster of galaxies.

Classical Cepheids are young, relatively massive giant stars ($M \geq
3~M_\odot$). They are excellent objects
for studying stellar pulsation mechanisms and testing stellar evolution
theory. Although pulsating and evolutionary status of Cepheids seemed to
be well established, both theories predicted contradictory masses
differing up to 20--30\%. This problem called often as ``Cepheid mass
problem'' (Neilson \etal 2011 and references therein) remained unsolved
for decades. This situation remained quite uncomfortable as the
properties and mechanisms of the basic astrophysical standard candle
should be well understood.

Only precise empirical mass determination of a classical Cepheid could
shed light on the Cepheid mass problem. Unfortunately, in spite of
considerable efforts none classical Cepheid has been found for years in
an eclipsing binary system. Such a wide binary system would allow
dynamical mass determination of Cepheid component with the accuracy
limited only by the accuracy of observations. On the other hand, large
dimensions of the putative binary system containing a Cepheid component
make the chance of observing eclipses for the Earth observer very slim. 

The breakthrough occurred in the last two decades when the Optical
Gravitational Lensing Experiment (OGLE) long-term large-scale sky
variability survey started releasing huge samples of classical Cepheids
in the Magellanic Clouds and presented several Cepheids revealing
simultaneously pulsating and eclipsing variability (Udalski \etal 1999,
Soszy{\'n}ski \etal 2008, 2010, 2012). These detections were followed by
an extensive spectroscopic campaign conducted by the Araucaria Project
(Gieren \etal 2005), which led to the first
determination of the classical Cepheid mass with the accuracy of the
order of one percent and final empirical word on the Cepheid mass
problem (Pietrzy{\'n}ski \etal 2010). A few more precise mass
determinations of the OGLE eclipsing Cepheids followed in the next years
(\eg Pietrzyñski \etal 2011, Pilecki \etal 2015, Gieren \etal 2015).

The final release of the OGLE collection of classical Cepheid in the
Magellanic System (Soszy{\'n}ski \etal 2015) contains almost ten
thousand of these pulsating stars located in the Large and Small Magellanic
Clouds and in the Magellanic Bridge connecting these two nearby
galaxies. This sample includes almost all classical Cepheids from this
System. It is the most uniform, and complete sample of classical
Cepheids in modern astrophysics with very precise photometry -- ideal
for extensive studies of Cepheids themselves, Magellanic System
structure, pulsating and evolutionary theories, etc.

The final search for classical Cepheids in the Magellanic System
conducted on the data collected during the fourth phase of the OGLE
survey (OGLE-IV, Udalski \etal 2015), led to the discovery of a
considerable sample of new interesting non-standard classical Cepheids.
Here we concentrate on classical Cepheids in eclipsing binary systems.
We present new candidates and also summarize all the OGLE past
discoveries of eclipsing binary systems with Cepheid component(s),
providing their status. Finally, we show empirically that the light-time
effect can be successfully used for selection of an additional sample of
Cepheids in binary systems -- those containing non-eclipsing objects.

\Section{Observations and Photometric Data}

Photometric data of classical Cepheids from the Magellanic System
presented in this paper come from the OGLE-IV survey 
(Udalski \etal 2015). They were collected in the years
2010--2015 with the 1.3-m Warsaw telescope equipped with the 32-CCD
detector mosaic camera covering 1.4 square degrees in the sky with the
scale of 0.26~arcsec/pixel. For technical details of the OGLE-IV
observing setup and strategy the reader is referred to Udalski \etal
(2015). In some cases we supplement OGLE-IV data with photometry
collected during earlier phases of the OGLE project.

The OGLE collection of classical Cepheids in the Magellanic System has
been recently presented by Soszy{\'n}ski \etal (2015). It contains almost
10\ 000 objects from the Large and Small Magellanic Clouds and
Magellanic Bridge carefully selected and well characterized. The sample
is very homogeneous and complete (>90\%). The main OGLE-IV survey of the
Magellanic System was conducted in the standard $I$-band supplemented with 
smaller number of standard $V$-band observations for color information.
Thus, the number of epochs in the $I$-band is considerably larger (up to
750) than in the $V$-band (up to 250). More details on the OGLE Cepheid
collection can be found in Soszy{\'n}ski \etal (2015).

\Section{Cepheids in Eclipsing Binary Systems}

Eclipsing binary systems hosting Cepheids have several interesting
astrophysical applications. First, photometric and spectroscopic
analyses of such objects may provide masses and radii of pulsating stars
with great accuracy. It is particularly important in the context of the
already mentioned longstanding mass discrepancy problem for classical
Cepheids. Second, detached double-lined eclipsing binary systems
containing Cepheids are promising tools to calibrate the Cepheid PL
relation, since a careful analysis of such systems may provide their
precise distances. Third, as shown by Pilecki \etal (2013) and Gieren
\etal (2015), it is possible to precisely determine the so called {\it
p}-factor -- the crucial parameter in the Baade-Wesselink method of
distance determinations to Cepheids.

Classical Cepheids that are members of eclipsing binary systems are very
rare objects. No such star is known in the Milky Way. All of the currently
known objects of this type belong to the Magellanic Clouds and all of them
were discovered by the OGLE survey.\footnote{All three pulsating stars with
eclipses discovered in the LMC by the MACHO project (Welch \etal 1999) were
type~II Cepheids.} The first ``eclipsing'' classical Cepheid
(OGLE-LMC-CEP-2532) was detected by Udalski \etal (1999) in the LMC. Then,
Soszyñski \etal (2008, 2010, 2012) reported six more Cepheids with
eclipsing modulation superimposed on the pulsation light curves. The binary
nature of these stars had to be confirmed spectroscopically.

OGLE-LMC-CEP-0227 was the first confirmed double-lined physical system
containing a classical Cepheid (Pietrzyñski \etal 2010). In that paper,
Pietrzyñski \etal (2010) measured the dynamical mass of
OGLE-LMC-CEP-0227 with an unprecedented accuracy of 1\% and the results
of this study ended the dispute about the predicted masses of Cepheids.
Other OGLE classical Cepheids in eclipsing binary systems were studied
in a similar way by Pietrzyñski \etal (2011), Gieren \etal (2014, 2015),
and Pilecki \etal (2013, 2015). Table~3 summarizes the confirmed and
candidate eclipsing binary systems containing classical Cepheids.

\MakeTableee{lccl}{12.5cm}{Classical Cepheids with eclipsing variability superimposed on the light curves.}
{\hline
\noalign{\vskip3pt}
\multicolumn{1}{c}{Identifier} & $P_{\rm puls}$ & $P_{\rm orb}$ & \multicolumn{1}{c}{Reference(s)} \\
                               & [days]         & [days]        &          \\
\noalign{\vskip3pt}
\hline
\noalign{\vskip3pt}
OGLE-LMC-CEP-0227 & 3.797078 & 309.67   & Soszyñski \etal (2008), \\
                  &          &          & Pietrzyñski \etal (2010), \\
                  &          &          & Pilecki \etal (2013) \\
OGLE-LMC-CEP-1718 & 2.480868 & 412.81   & Soszyñski \etal (2008), \\
                  & 1.963655 &          & Gieren \etal (2014) \\
OGLE-LMC-CEP-1812 & 1.312904 & 551.80   & Soszyñski \etal (2008), \\
                  &          &          & Pietrzyñski \etal (2011) \\
OGLE-LMC-CEP-2532 & 2.035352 & 800.42   & Udalski (1999), \\
                  &          &          & Soszyñski \etal (2008), \\
                  &          &          & Pilecki \etal (2015) \\
OGLE-LMC-CEP-3782 & 4.506933 & 242.72   & this paper \\
OGLE-LMC-CEP-4506 & 2.987824 & 1550.4   & Soszyñski \etal (2012), \\
                  &          &          & Gieren \etal (2015) \\
OGLE-SMC-CEP-0411 & 1.101093 & ~~43.50  & Soszyñski \etal (2010) \\
OGLE-SMC-CEP-1996 & 2.317943 & ~~95.60  & Soszyñski \etal (2010) \\
OGLE-SMC-CEP-2199 & 3.373778 & 1210.6   & this paper \\
OGLE-SMC-CEP-3235 & 0.864109 & $\sim$4200 & this paper \\
OGLE-SMC-CEP-4795 & 3.150119 & ~~202.6  & this paper \\
\noalign{\vskip3pt}
\hline}

\Section{New Eclipsing Cepheids}

In this paper, we report the discovery of additional four Cepheids
revealing eclipses superimposed on the regular standard pulsation
variability. All were found during our analysis of the OGLE-IV
Cepheid photometry (Soszy{\'n}ski \etal 2015).

In three cases we observed more than one eclipse. Thus, it was possible
to determine the orbital periods. Light curves of these stars are shown
in Fig.~1. It should be noted that in all these three cases eclipses are
rather shallow. This may suggest either grazing eclipses or
alternatively, fake objects -- unresolved blends of a Cepheid and an
unrelated eclipsing binary system in the seeing disk. However, we should
also note that the derived orbital periods are rather long (from over
200 days to over one thousand), thus, the eclipsing systems must be wide
and the latter -- blend -- possibility is not very likely.   

\begin{figure}[t]
\begin{center}
\includegraphics[width=10cm]{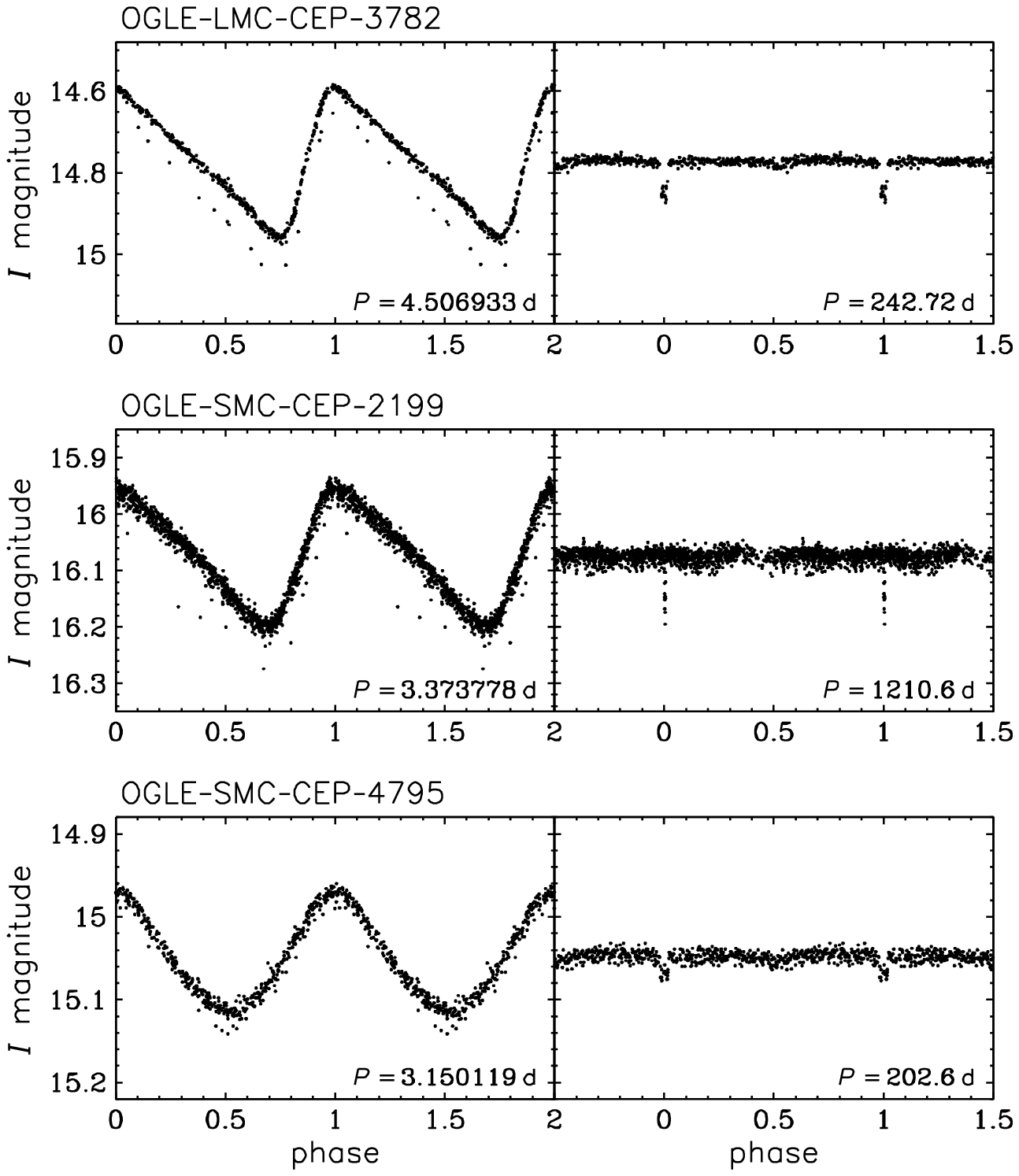}
\end{center}
\FigCap{Light curves of Cepheids with additional eclipsing variability.
{\it Left panels} show the original photometric data folded with the
pulsation periods. {\it Right panels} show the eclipsing light curves
after subtracting the Cepheid component. The ranges of magnitudes are the
same in each pair of the panels.}
\end{figure}

\begin{figure}[t]
\includegraphics[width=12.7cm]{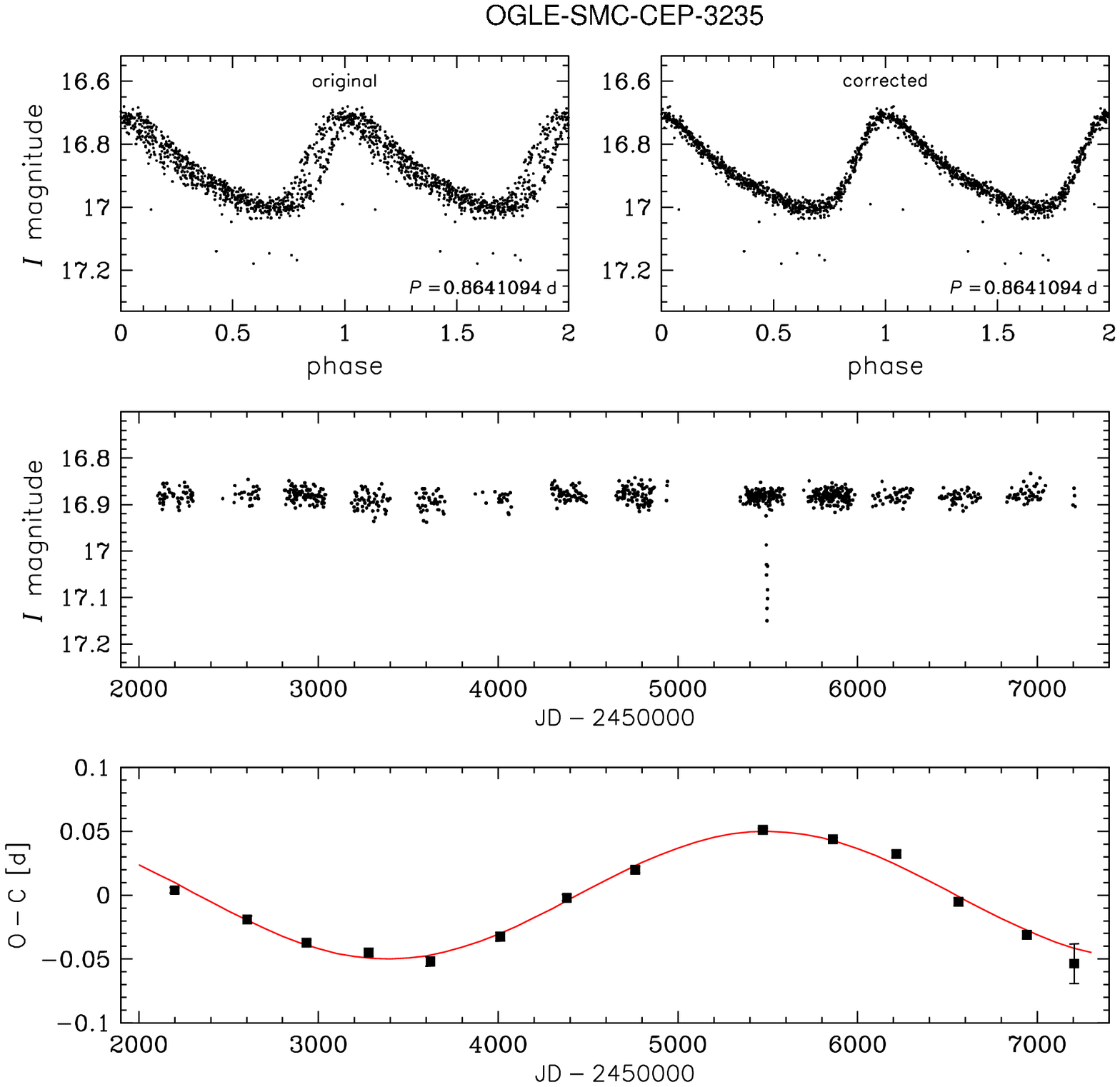}
\FigCap{OGLE-SMC-CEP-3235 -- a classical Cepheid in an eclipsing binary
system. {\it Upper left panel} shows the original light curve folded with
the pulsation period, while the {\it upper right panel} displays the same light
curve corrected for the light-time effect. {\it Middle panel} presents the
light curve prewhitened with the pulsations. {\it Lower panel} shows the
$O-C$ diagram for OGLE-SMC-CEP-3235 obtained in the years 2001-2015.}
\end{figure}

The fourth case -- OGLE-SMC-CEP-3235 -- requires special attention
because this first-overtone Cepheid experienced only one deep eclipse
during the 15-year time span of the OGLE-III and OGLE-IV surveys
(Fig.~2). However, we noticed that the pulsation light curve of
OGLE-SMC-CEP-3235 cannot be perfectly phased with a constant period.
The observed minus calculated ($O-C$) diagram computed for this light
curve revealed long-term sinusoidal-like variations of the pulsation
phases which are probably associated with the light-time effect caused
by the orbital motion of the Cepheid. 

Thus, we adjusted the orbital and pulsation periods by the minimization
of the point scatter in the light curve corrected for the $O-C$ diagram
(assuming its sinusoidal shape, \ie a circular orbit of the system). The
light curve corrected for the light-time effect is shown in the upper
right panel of Fig.~2, while the resulting $O-C$ diagram is displayed in
the lower panel.

Analysis of the $O-C$ diagram provides some information about the
system. The orbital period is $4200\pm300$~d. This value is quite
uncertain because the adjusted orbital period was strongly correlated
with the true pulsation period which was a free parameter in our
adjustment procedure. The precise value of the orbital (and pulsation)
period will be known when the next eclipse will be recorded or longer
span variability in the $O-C$ diagram will be collected.

The $O-C$ diagram reaches a maximum during the eclipse, which indicates
that the Cepheid was eclipsed by its binary companion. Unfortunately, the
secondary eclipse has not been caught during the  OGLE-III and OGLE-IV
monitoring. This may suggest that either the companion has a much lower
temperature than the Cepheid or, simply, that the secondary eclipse
occurred during a gap between the observing seasons. The amplitude of
the $O-C$ variations ($0.050\pm0.005$~d) can be easily transformed into
the semi-major axis of the Cepheid orbit, equal to $8.7\pm0.9$~AU.

\Section{Light-Time Effect in Binary Cepheids}

Encouraged by our analysis of OGLE-SMC-CEP-3235 where the light-time
effect seems to be very well pronounced and well determined we decided
to perform a similar search for the light-time effect in other known
binary eclipsing systems containing a Cepheid component. Such a test was
supposed to answer the question if the light-time effect can be a useful
tool for searching Cepheids in non-eclipsing binary systems. It is well
known that the $O-C$ diagrams of classical Cepheids reveal long-term
variability which can be mistakenly interpreted as the light-time effect
(Poleski 2008).

First, we note that Pilecki \etal (2013) have already found a marginal
light-time effect in the first eclipsing Cepheid -- OGLE-LMC-CEP-0227.
Then we analyzed the remaining three confirmed Cepheid eclipsing
binaries. We quickly realized that the most promising objects for
positive detection of the light-time effect are only those Cepheids that
reveal steep brightness rise with a large sensitivity to phase shifts,
\ie those with a short pulsation period. Additional obvious conditions are
large orbit making the amplitude of the light-time effect larger and low
eccentricity making the $O-C$ changes sinusoidal-like. Thus, only
OGLE-LMC-CEP-1812 remained the most promising object for our test.

\begin{figure}[t]
\includegraphics[width=12.7cm]{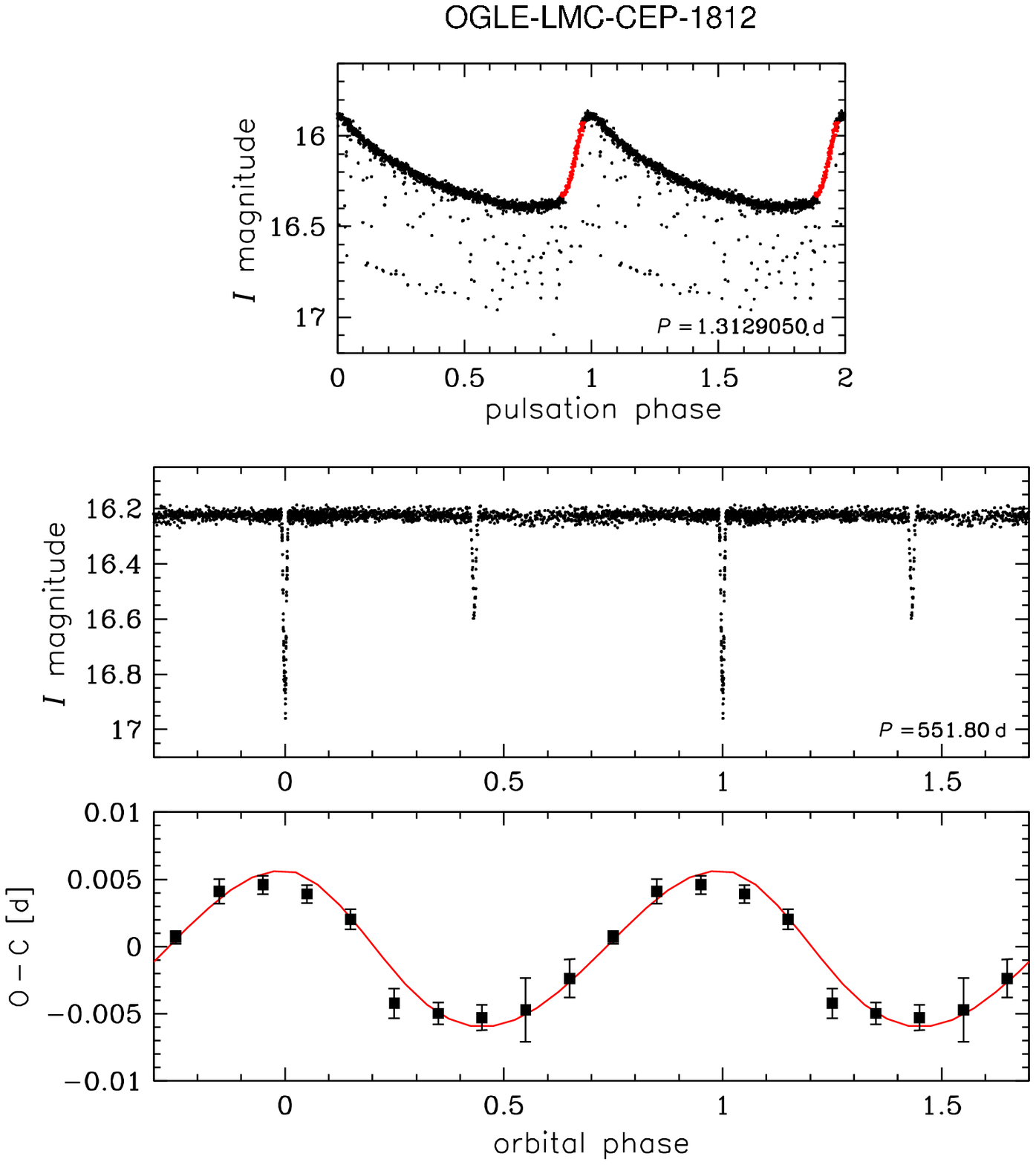}
\FigCap{OGLE-LMC-CEP-1812 -- a classical Cepheid in an eclipsing binary
system. {\it Upper panel} shows the original light curve folded with the
pulsation period. Red points indicate observations used for the
construction of the $O-C$ diagram. {\it Middle panel} presents the light curve after
removing the Cepheid pulsation, folded with the eclipse ephemeris. {\it
Lower panel} shows the averaged $O-C$ diagram for OGLE-LMC-CEP-1812
folded with the eclipse ephemeris. Red curve shows the $O-C$ diagram
obtained from the spectroscopic observations by Pietrzyñski \etal (2011).}
\end{figure}

Fig.~3 shows results of the search for light-time effect of
OGLE-LMC-CEP-1812 in the OGLE-III and OGLE-IV data. To minimize the noise,
only observations obtained between pulsating phases 0.88--0.98 (\ie during the rapid rise to maximum light) were used for the
shift determination. They are the most sensitive for phase (period)
changes. The upper panel of Fig.~3 presents the light curve of
OGLE-LMC-CEP-1812 folded with the pulsation period while the middle one
-- eclipsing light curve of this system after removing the pulsation
component and folded with the eclipsing ephemeris.

The lower panel shows the $O-C$ obtained during the period 2001--2015
with individual data averaged in bins season-wide and also folded with
the eclipsing ephemeris (orbital period). The nearly sinusoidal variability of
the $O-C$ is clearly detectable. Its period is consistent with the
eclipsing period of the system and the eclipses coincide with the
extrema of the $O-C$ variability confirming that the variability is
caused by the real light-time effect. The amplitude of the $O-C$
variation is also consistent with the expected magnitude of this effect
calculated based on the real system parameters (orbit size) as
determined by Pietrzyñski \etal (2011) -- red curve.

Positive detection of the light-time effect in two eclipsing binaries
containing a Cepheid component clearly indicates that the method can
also be extended to searching for Cepheids in non-eclipsing binaries. The
long span of the OGLE observations reaching 15--20 years should allow
detections of even very wide systems. One has to be, however, careful to
avoid over-interpretation of the long-term $O-C$ variability. Contrary
to the presented here cases of light-time effect, additional
information confirming its reality -- eclipses -- will not be, of
course, available.  Nevertheless, confirmation of binarity can still be
possible with spectroscopic follow-up observations. Additionally, it may
be necessary to include in some cases possible long-term change of the
Cepheid pulsating period. 

\begin{figure}[t]
\includegraphics[width=12.7cm]{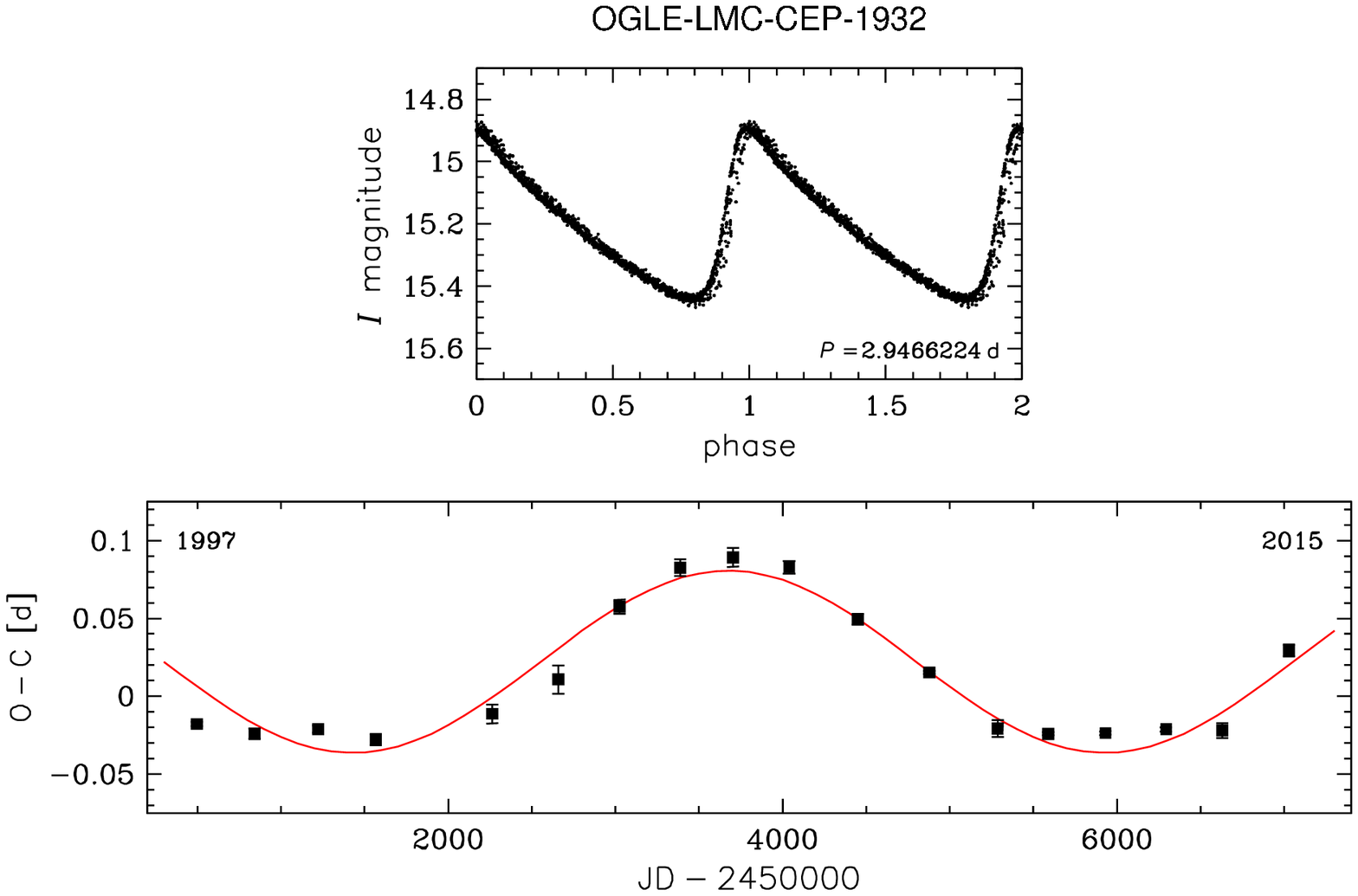}
\FigCap{OGLE-LMC-CEP-1932 -- candidate for a non-eclipsing binary system
with a classical Cepheid component. {\it Upper panel} shows the original
light curve folded with the mean pulsation period. Note the
scatter in the rising branch of the light curve. This part is the most
sensitive to the light-time effect. {\it Lower panel} shows the averaged
$O-C$ diagram for OGLE-LMC-CEP-1932 obtained in the years 1997-2015. Red
sinusoid is a crude fit to the $O-C$ data indicating a possible orbital
period of the order of 4500~d.}
\end{figure}

Careful analysis of the entire OGLE collection of Cepheids in the
Magellanic System should allow statistical studies of the binarity of
Cepheids. Such a study is underway. Fig.~4 presents a potentially
promising candidate OGLE-LMC-CEP-1932. It was found by Poleski (2008)
and further observations during OGLE-III and OGLE-IV phases indicate
that the $O-C$ diagram may show periodic variability with a period of
about 4500~d. Although the simple sinusoidal fit presented in Fig.~4
is far from being acceptable, this is a clear example that the detection
of non-eclipsing binaries with classical Cepheid component can be
feasible with the OGLE photometry. It should be also noted that a similar
study of the OGLE-detected RR Lyrae pulsators has been recently
conducted by Hajdu \etal (2015) leading to the detection of several
non-eclipsing binary candidates containing an RR-Lyrae component.
  
\Acknow{The OGLE project has received funding from the Polish
National Science Centre grant MAESTRO no. 2014/14/A/ST9/00121 to AU.
This work has been supported by the Polish National Science
Centre grant no. DEC-2011/03/B/ST9/02573.}

\end{document}